\documentclass[aps,prd,reprint,superscriptaddress]{revtex4-1}

\usepackage[utf8]{inputenc} 
\usepackage{physics}
\usepackage[english]{babel}
\usepackage{graphicx,color,overpic,mathtools}
\usepackage{amsthm,amsmath,amssymb,hyperref}
\usepackage{physics}
\graphicspath{ {images/} }
\addto\captionsspanish{}
\hypersetup{
    colorlinks=false,
    pdfborder={0 0 0},
}

\begin{document}

\title{Schwarzschild geometry counterpart in semiclassical gravity}
\author{Julio Arrechea}
\email{arrechea@iaa.es}
\affiliation{Instituto de Astrof\'isica de Andaluc\'ia (IAA-CSIC),
Glorieta de la Astronom\'ia, 18008 Granada, Spain}
\author{Carlos Barcel\'o} 
\email{carlos@iaa.es}
\affiliation{Instituto de Astrof\'isica de Andaluc\'ia (IAA-CSIC),
Glorieta de la Astronom\'ia, 18008 Granada, Spain}
\author{Ra\'ul Carballo-Rubio}
\email{raul.carballorubio@ucf.edu}
\affiliation{Florida Space Institute, 12354 Research Parkway, Partnership 1, Orlando, FL 32826-0650, USA}
\affiliation{IFPU, Institute for Fundamental Physics of the Universe, Via Beirut 2, 34014 Trieste, Italy}
\author{Luis J. Garay} 
\email{luisj.garay@ucm.es}
\affiliation{Departamento de F\'{\i}sica Te\'orica and IPARCOS, Universidad Complutense de Madrid, 28040 Madrid, España}  
\affiliation{Instituto de Estructura de la Materia (IEM-CSIC), Serrano 121, 28006 Madrid, Spain}

\begin{abstract}
We investigate the effects of vacuum polarization on vacuum static spherically-symmetric spacetimes. We start from the Polyakov approximation to the renormalized stress-energy tensor (RSET) of a minimally coupled massless scalar field. This RSET is not regular at $r=0$, so we define a regularized version of the Polyakov RSET. Using this Regularized RSET, and under the previous symmetry assumptions, we find all the solutions to the semiclassical field equations in vacuum. The resulting counterpart to the Schwarzschild classical geometry substitutes the presence of an event horizon by a wormhole throat that connects an external asymptotically flat region with an internal asymptotic region possessing a naked singularity: there are no semiclassical vacuum solutions with well-defined Cauchy surfaces. We also show that the Regularized Polyakov RSET allows for wormhole geometries of arbitrarily small throat radius. This analysis paves the way to future investigations of proper stellar configurations with an internal non-vacuum region.
\end{abstract}

\maketitle

\section{Introduction}

The presence of curvature in a spacetime makes it impossible to completely subtract  the zero-point contribution of the quantum fields living on that spacetime. For this reason, the renormalized expectation value of the stress-energy-tensor operator of the quantum fields (the RSET for short) must be taken into account as an additional source of gravity. Hence, strictly speaking, in the presence of curvature, even regions of spacetime in which there is no classical matter are no longer empty, but filled with an effective semiclassical substance. Understanding the effects of the semiclassical contributions to the Einstein equations is the subject matter of semiclassical gravity (see~\cite{birrell1984} for instance). It is reasonable to expect that such quantum deviations should be derivable from any acceptable theory of quantum gravity (in fact, this is arguably a minimal requirement to be satisfied by any such theory). In this sense, one can argue that semiclassical effects are our most robust window into the first quantum deviations from the classical theory due to gravity itself. In addition, one would expect that these first modifications become relevant in extreme gravitational scenarios, containing either singularities or horizons.

Here we are interested in semiclassical effects in stellar-like configurations, specifically in static and spherically symmetric spacetimes having a single asymptotically flat region. It is well known that under some mild conditions there only exist regular stellar configurations when the compactness $C$ of the star, defined as the ratio between its gravitational radius $2M$ (twice the mass of the star) and its actual radius $R$, is smaller than the so-called Buchdahl limit: $C<8/9$~\cite{Buchdahl1959,Andreasson2007,Karageorgis2007}. For these regular sub-Buchdahl stars, semiclassical effects are expected to be negligible throughout their structure. However, remarkably and as far as we know, there is no complete formal proof that this is the case; instead one reaches this reasonable conclusion by putting together a series of separate arguments. Essentially, the RSET is always multiplied by the square of Planck length (proportional to $\hbar$) and in sub-Buchdahl stars there is no reason to expect that this small number is being compensated by vacuum polarization effects. Nonetheless, one must keep in mind that there does not exist an exact calculation of the RSET for a normal (sub-Buchdahl) star where one can explicitly check the smallness of semiclassical effects. Two approaches to this issue are the approximate schemes presented in \cite{Hiscock1987} and \cite{Satz2005}. In the first work the RSET is constructed from local tensors and then computed over the background of a uniform density star. The latter obtains vacuum polarization contributions in the weak-field limit by means of a non-local approximation. For Newtonian stars, non-local contributions coming from the RSET are almost constant throughout the structure and are seen to dominate everywhere over local terms. These works involve computations over a fixed background and therefore do not explore in a self-consistent manner how vacuum polarization modifies the background metric.

The reason for the described situation is that the most easily self-consistently tractable approximations to the RSET diverge at the center of regular stars, thus making it difficult to check the smallness of the RSET near the center. For instance, a RSET obtained via the so-called Polyakov approximation~\cite{DaviesFulling1977, Polyakov1981, Balbinotetal2000} is divergent due to the $1/4\pi r^2$ factor that is necessary to have a conserved 4-dimensional tensor. In the case of an $s$-wave-approximation RSET (e.g.~\cite{FabbriNavarro2005,Fabbrietal2005}), which should provide a better approximation, one would expect finite components at the radial origin. However, perhaps surprisingly, this is not the case, at least using the expressions obtained in~\cite{FabbriNavarro2005}, which are still divergent at $r=0$. The most sensible conclusion is that the expressions obtained in~\cite{FabbriNavarro2005,Fabbrietal2005} are not appropriate to deal with regular stars, since the modes with respect to which the quantization of the scalar field is performed are suited to describe black hole or wormhole spacetimes, with either a singularity at $r=0$ or without such $r=0$ point at all. For regular stars one should impose sensible boundary conditions at $r=0$, which are not contemplated in~\cite{FabbriNavarro2005,Fabbrietal2005}. While more refined approximations to the RSET  and
the exact RSET itself (see \cite{Andersonetal1995}) should be regular at $r=0$ for sub-Buchdahl stars, the complexity involved in dealing with these tensors hinders this analysis which, moreover, only considers the Hartle-Hawking vacuum state. 

Even though one does not expect any surprises for sub-Buchdahl stars, the problem of the finiteness of the RSET at $r=0$ becomes more pressing when trying to analyze semiclassical effects for classical configurations that are irregular themselves (super-Buchdahl stellar configurations, with $C \geq 8/9$)~\cite{Carballo-Rubio2017,HoMatsuo2017} which is, in fact, the actual central motivation for the present work. In order to establish the regularity of these configurations within the semiclassical approximation, it is necessary to find a framework that does not present the issue of the divergence of the approximations to the RSET used in these works. Thus, the problem of knowing whether non-singular semiclassical solutions exist or not becomes entangled with the behavior of the relevant geometrical functions at $r=0$.

Here, as a warm up study of the effect of the RSET in regions close to zero radius, we are going to obtain and analyze the complete set of self-consistent vacuum solutions (with no classical matter) of the semiclassical theory using the Polyakov RSET modified by a suitable regulator that ensures finiteness at $r=0$. We will analyze the effect of this regulator in the set of solutions, extending the family of solutions obtained in~\cite{Fabbrietal2005,HoMatsuo2017}, where no regulator was used.  In our analysis we are going to consider the field to be in the Boulware vacuum state, the natural vacuum state if one is looking for genuine static and asymptotically flat spacetimes. The expressions for the Polyakov RSET in the Boulware vacuum can be obtained just by turning off the energy fluxes in the expressions from \cite{DaviesFulling1977}, which in a fixed background calculation leads to a divergent behaviour at the event horizon of a Schwarzschild black hole. In a self-consistent treatment, this divergence does not appear and just indicates where huge backreaction effects are going to kick in, possibly modifying the geometry non-perturbatively. In this work we shall not analyze dynamical situations associated with the Unruh vacuum state, nor static situations corresponding to a Hartle-Hawking vacuum state. This last state is suitable for the description of black holes in equilibrium with a thermal bath, which in background calculations results instead in finite semiclassical contributions at the event horizon, at least for Polyakov and $s$-wave approximations. Thermal equilibrium could be attained for example by putting the black hole inside a reflecting box.

In the analysis that follows we find that the Schwarzschild geometry counterpart in semiclassical
gravity is a non-symmetric wormhole geometry with an asymptotically flat region and a singular internal asymptotic region at a finite proper radial distance, which constitutes a naked singularity. The size of the
throat can be made arbitrarily small by making small the asymptotic ADM mass of the configuration. In addition, our analysis provides a rigorous proof of the form and uniqueness of the obtained solutions.

In the next section we will start preparing the way for the main analyses in the paper. Section~\ref{Sec:Regularized} describes the characteristics of the regular RSET we are considering. 
Then, we shall pass to the main part (sections \ref{Sec:Self-consistent} and \ref{Sec:Solution}) of the paper in which we obtain all the self-consistent vacuum solutions of our regularized semiclassical theory. Section \ref{Sec: Discussion} will be devoted to discussing the main characteristics of those solutions. Finally, we will summarize our findings and point out some future points to address.  We will work in units $G=c=1$, and for convenience we will make use of a rescaled Planck length $l_{\rm P}^{2}\equiv\hbar/\sqrt{12\pi}$.

\section{Preliminaries}
\label{Sec:Preliminars}
 
It is well known that spherically-symmetric vacuum solutions in classical general relativity are described by the Schwarzschild family, parameterized by the ADM mass of the geometry. In this paper, we study which geometries take the role of Schwarzschild geometry when the semiclassical effects of quantum vacuum polarization are taken into account. As we will explicitly show in the discussion below, solutions resulting from our analysis cannot have non-extremal trapping horizons, i.e. with a nonvanishing surface gravity (being the geometries static and spherically symmetric, we can use indistinguishably the names trapping/apparent/event). As we will see, no horizon of any kind (extremal or not) shows up in our analysis, so we
can start by writing down a sufficiently general line element as
\begin{equation}\label{eq:2Dspacetime}
     ds^{2}=-e^{2\phi(r)}dt^{2}+\frac{1}{1-C(r)}dr^{2}+r^{2}d\Omega^{2}.
\end{equation}
There are two distinct notions of compactness relevant for our analysis. The function $\phi(r)$ can be thought of as encoding a redshift compactness, while the function $C(r)$ resembles the energy compactness, which, in the context of relativistic stars, provides a notion of the amount of energy density contained inside a sphere of radius $r$. In fact, this function can be written as $C(r)=2m(r)/r$, where $m(r)$ is the Misner-Sharp mass~(e.g. \cite{Misner1964,Hernandez1966,Hayward1994}).

\subsection{The classical vacuum solution}

Using~(\ref{eq:2Dspacetime}) the classical vacuum Einstein equations have the form:
\begin{align}\label{eq:ClassicalEqs}
G_{tt}=
&
C+rC'=0\nonumber,\\
G_{rr}=
&
C(1+2r\psi)-2r\psi=0,\\
G_{\theta\theta}=
\frac{G_{\varphi\varphi}}{\sin^{2}{\theta}}=
&
2r\left(1-C\right)\big[\psi'+\psi^{2}+\frac{\psi}{r} \nonumber \vphantom{\psi'+\psi\left(\psi+\frac{1}{r}\right)}+C'(\psi+r)\big]=0\nonumber.
\end{align}
Here $\psi=\phi'$ with $'$ denoting derivatives with respect to the radial coordinate. Owing to Bianchi identities, solutions can be uniquely determined by use of the $tt$ and $rr$ equations, the angular components being a consequence of the first two. It is straightforward to realize how the Schwarzschild family of solutions is recovered in these coordinates for $r>2M$:
\begin{equation}\label{eq:ClassicalSchwarzschild}
\phi(r)=\frac{1}{2}\ln \left(1-\frac{2M}{r}\right)+\phi_{0},~~~~C(r)=\frac{2M}{r}.
\end{equation}
Here, $\phi_{0}$ represents just an unobservable rescaling of time. Note that in vacuum the time component of the metric is the inverse of the radial component. In this case the notions of red-shift and energy compactness coincide. As is well known, this gives place to the presence of an event horizon at $r_{\rm S}=2M$ and to the extendibility of the geometry beyond the horizon, which can be seen by using other sets of coordinates, such as Kruskal-Szekeres \cite{Kruskal1959,Szekeres1960}.

In the semiclassical theory vacuum energy acts as a matter source so that the right-hand side of \eqref{eq:ClassicalEqs} is not equal to zero (as long as spacetime is not strictly flat). As a consequence, redshift and energy compactness become distinct notions. This differentiation, as we will discuss in detail below, makes remarkably different the family of solutions that in the semiclassical theory plays the role of the Schwarzschild family. As a side note, let us mention that such differentiation takes place every time a matter source is introduced and it is independent of whether its origin is classical or quantum; for instance, this is the case for dirty black holes~\cite{Visser1992}
or equivalently, dirty stellar configurations.
 
 \subsection{Vacuum semiclassical gravity in the Polyakov approximation}

The zero-point energy of quantum fields acts as a source of spacetime curvature, and this relation is given by the semiclassical Einstein equations:
\begin{equation}\label{eq:SemiclassicalEinstein}
G_{\mu\nu}=8\pi\hbar\langle\hat{T}_{\mu\nu}\rangle.
\end{equation}
In the following we analyze the simplest scenario of having a single quantum massless scalar field (although all our results are, in the approximation that we will be using, equally valid for an arbitrary number of scalar and fermion fields, as long as these are massless). The expectation value of its RSET is taken in the Boulware vacuum: the natural vacuum state for static situations.

For a purely $(1+1)$-dimensional geometry, the scalar field equation of motion becomes conformally invariant, allowing to find an exact expression for a conserved RSET~\cite{DaviesFulling1977}. This is the so-called 2-dimensional Polyakov RSET $ \langle{\hat{T}_{\mu\nu}}\rangle^{\rm{P}2}$ \cite{FabbriNavarro2005}. Taking this $1+1$ geometry to be the $(t,r)$ sector in Eq.~(\ref{eq:2Dspacetime}), the components of this RSET are:
\begin{align}\label{eq:RSETcomponents}
        \langle{\hat{T}_{rr}}\rangle^{\rm{P}2}=
        &
        -\frac{l_{\rm P}^{2}\psi^{2}}{2},\quad \quad \langle{\hat{T}_{tr}}\rangle^{\rm{P}2}=\langle{\hat{T}_{rt}}\rangle^{\rm{P}2}=0,\nonumber \\
        \langle{\hat{T}_{tt}}\rangle^{\rm{P}2}=
        &
        \frac{l_{\rm P}^{2}e^{2\phi}}{2}\left[2\psi'(1-C)+\psi^{2}(1-C)-\psi C'\right].
\end{align}

Our goal is to compute semiclassical contributions in a realistic (3+1) setting though. From the tensor \eqref{eq:RSETcomponents}, conserved in (1+1) dimensions by construction, we can build a (3+1) tensor which is now conserved in (3+1) dimensions:
\begin{equation}\label{eq:DimTransf}
        \langle{\hat{T}_{\mu\nu}}\rangle^{\rm{P}4}=\frac{1}{4\pi r^{2}}\delta^a_\mu\delta^b_\nu\langle{\hat{T}_{ab}}\rangle^{\rm{P}2}.
\end{equation}
In this expression, as well as in the rest of the paper, Greek indices take four values, while Latin indices take only two: $r$ and $t$. The multiplicative factor $1/4\pi r^2$ ensures conservation of $\langle{\hat{T}_{\mu\nu}}\rangle^{\rm{P}4}$. In the following we will eliminate the number 4 in our notation as all of our discussions will take place in $3+1$ dimensions.

In the context of the $3+1$ theory, this Polyakov RSET can be obtained by taking two approximations in the equations of motion of the scalar field. Firstly, the field admits a decomposition in spherical harmonics, from which only the $s$-wave component is considered. There are indications that higher multipoles provide subdominant contributions to the exact RSET when compared to the $s$-wave contribution \cite{Sanchez1978}. There have been attempts to compute the RSET including arbitrarily-high multipoles \cite{Andersonetal1995}.  However, the intricacy of the resulting expressions makes it hard to treat them self-consistently. Secondly, the other approximation invoked consists in neglecting the potential in the equation of motion for the $s$-wave component. In doing so, the modes of the field are not subject to backscattering, meaning that the outgoing and ingoing mode contributions are decoupled. The simplicity the construction gains within this double approximation is well worth the loss of accuracy regarding the information content of the RSET, at least for many applications. 

\subsection{Regularity at $r=0$}

The Polyakov RSET is well suited to qualitatively account for the behavior of the exact RSET throughout a spherically symmetric spacetime, only if $r=0$ is not approached. As we are going to see in the next subsection, the Polyakov RSET diverges at $r=0$ even in geometries which are regular at their center. Essentially the reason for that is that the absence of backscattering causes any ingoing or outgoing shell-like wave to concentrate its energy at the radial origin. If backscattering were taken into account, the modes would be smeared out and the central singularity would be excised from the RSET. However, in the $s$-wave treatment carried out in Ref. \cite{FabbriNavarro2005} a divergent behavior at $r=0$ is still present, making us suspect the inappropriateness of these expressions to deal with regular stars. We think that the problem comes from the absence of a reflective boundary condition on the modes at $r=0$. Notice that this does not invalidate the results in \cite{FabbriNavarro2005} as these authors apply these expressions to wormhole and black hole-like configurations, where $r=0$ is either non-existent or non-reflective.

For the metrics we are analyzing, the Kretschmann scalar $\mathcal{K}=R_{\mu\nu\rho\sigma}R^{\mu\nu\rho\sigma}$ is 
\begin{align}
     \mathcal{K}=
&
\frac{4C^{2}}{r^{4}}+\frac{2C'^{2}}{r^{2}}+\frac{8\psi^{2}(1-C)^{2} }{r^{2}}\nonumber\\
&
     +\left[\psi C'-2(\psi^{2}+\psi')(1-C)\right]^{2},
\end{align}
thus being a positive definite quantity. Therefore, a finite Kretschmann scalar ensures that the geometry is devoid of any curvature singularity constructed from the Riemann tensor.
 
By means of this expression we can derive the conditions that the metric functions need to satisfy at $r=0$ in order to guarantee a finite $\mathcal{K}$. Regularity implies that the compactness $C$ must vanish at least quadratically in $r$, while $\psi$ must vanish at least linearly. Written in terms of the temporal and spatial components of the metric $g_{tt}$ and $g_{rr}$, these conditions require that:
\begin{equation}\label{eq:RegCond}
     -e^{2\phi}\simeq\beta+\gamma r^{2}+\mathcal{O}(r^{3}),\quad (1-C)^{-1}\simeq1+\kappa r^{2}+\mathcal{O}(r^{3}
     ),
\end{equation}
where $\beta,\gamma$ and $\kappa$ are constants. 

Given these geometries with a regular local behavior at $r=0$, we can prove that the Polyakov RSET is divergent at the origin. Indeed, by taking the first term in the $tt$ component of the Polyakov RSET,
\begin{equation}\label{eq:}
\langle \hat{T}_{tt}\rangle^{\rm P}= \frac{l_P^2 e^{2\phi}}{8\pi r^2} [2\psi'(1-C)+ \cdots~],
\end{equation}
we can see the existence of a $1/r^{2}$ divergence for the previously described behaviors $\psi\propto r$ and $C\propto r^{2}$.
Therefore, in the case that one proved the non-existence of regular
semiclassical solutions sourced by the Polyakov RSET this would be more
a proof of the inappropriateness of this RSET than of the non-existence
of self-consistent semiclassical solutions. Consequently, we explicitly
see that the Polyakov RSET is not suitable for the search of regular
self-consistent semiclassical stellar configurations.

In summary, seeking for balance between the tractability of the backreaction problem while avoiding the problem of the divergence at the origin is the central motivation for the following discussion.

\section{Regularized 3+1 Polyakov approximation}
\label{Sec:Regularized}
 
We have seen that the regularity of curvature invariants given by the conditions in Eq. \eqref{eq:RegCond} does not guarantee that the Polyakov RSET is regular as well; hence, the Polyakov RSET is not appropriate for the search of self-consistent solutions all the way down to $r=0$.

One possibility in order to obtain an appropriate RSET which is at least qualitatively trustable through the whole geometry is to regularize the Polyakov RSET. That the Polyakov RSET must be regularized in order to deal with practical situations has been noticed before, for example in the numerical implementation by Parentani and Piran~\cite{ParentaniPiran1994} of a semiclassical gravitational collapse. 

Following these authors, we introduce a cutoff in the $(t,r)$ sector of the Polyakov RSET as
\begin{equation}\label{eq:DistPol}
        \langle{\hat{T}_{ab}}\rangle^{\rm DP}\to \frac{4\pi r^{2}}{4\pi \left(r^{2}+\alpha l_{\rm P}^{2}\right)}\langle{\hat{T}_{ab}}\rangle^{\rm P},
\end{equation}
where taking $\alpha>0$ is sufficient to make this Distorted Polyakov RSET regular at $r=0$. However, it is straightforward to check that this regularization of the Polyakov RSET carries along the non-conservation of this object. Thus, finding a proper RSET which is both regular and conserved requires adding to the Distorted Polyakov RSET an additional compensatory piece. This compensatory term $\langle \hat{T}_{\mu\nu}\rangle^{\rm C}$ will be assumed to have only angular contributions.

The components of $\langle \hat{T}_{ab}\rangle^{\rm C}$ are obtained by taking the divergence of the total tensor:
\begin{equation}\label{eq:conservation}
    \nabla^{\mu}\left(\langle \hat{T}_{\mu\nu}\rangle^{\rm DP}+\langle \hat{T}_{\mu\nu}\rangle^{\rm C}\right)=0,
\end{equation}
so that their form depends on the multiplicative factor that permits us to go from $\langle{\hat{T}_{\mu\nu}}\rangle^{ \rm P }$ to $\langle{\hat{T}_{\mu\nu}}\rangle^{ \rm DP}$. This multiplicative factor constitutes an attempt to regularize the RSET in the most simple and mild way. As a result we obtain the Regularized Polyakov RSET:
\begin{equation}
    T^{\rm RP}_{\mu\nu}\equiv T^{\rm DP}_{\mu\nu}+T^{\rm C}_{\mu\nu},
\end{equation}
which is the RSET we are going to use in the rest of the paper.

Contrarily to the Polyakov RSET, the Regularized Polyakov RSET is regular at $r=0$ and contains non-vanishing angular components, both being features that a potentially regular $s$-wave RSET would share. Equation \eqref{eq:conservation} can be solved algebraically in order to show that the nonzero angular components of the Regularized Polyakov RSET are:
\begin{equation}
    \langle \hat{T}_{\theta\theta}\rangle^{\rm RP}=\frac{\langle \hat{T}_{\varphi\varphi}\rangle^{\rm RP}}{\sin^{2}\theta}=-\frac{\alpha r^2}{8\pi\left(\alpha+r^{2}/l_{\rm P}^{2}\right)^{2}}\psi^{2}(1-C).
\end{equation}
These components vanish when $\alpha=0$ and behave properly in the limit $r\to0$ when $\alpha\neq0$, by virtue of the regularity conditions (\ref{eq:RegCond}):
\begin{equation}
    \langle \hat{T}_{\theta\theta}(r\to0)\rangle^{\rm RP}\simeq-\frac{\gamma^{2} r^{4}}{8\pi\alpha\beta^{2}}+\mathcal{O}(r^{6}).
\end{equation}

\section{Self-consistent vacuum semiclassical equations}
\label{Sec:Self-consistent}
In what follows we will solve the semiclassical Einstein equations in vacuum sourced by the Regularized Polyakov RSET. We will write down the resulting system of equations in a simplified form to provide insight about the characteristics of the solutions.
The $rr$ and $tt$ components of the Einstein equations are, respectively:
\begin{align}\label{eq:srr}
    C=
    &
    \frac{2r \psi+l_{\rm P}^{2} r^{2} \psi^{2}/\left(r^{2}+\alpha l_{\rm P}^{2}\right)}{1+2r \psi+l_{\rm P}^{2} r^{2} \psi^{2}/\left(r^{2}+\alpha l_{\rm P}^{2}\right)},\\
    \label{eq:stt}
    C'=
    &
    \frac{-{C}/{r}
    +{r l_{\rm P}^{2}(1-C)}\left(
        \psi^{2}
        +2\psi'\right)/{(r^{2}+
    \alpha l_{\rm P}^{2})}
    }{1+{r l_{\rm P}^{2}}\psi/{(r^{2}+\alpha l_{\rm P}^{2})}}.
\end{align}
We can replace \eqref{eq:stt} by a first order differential equation for $\psi$. This is done by substituting (\ref{eq:srr}) and its first derivative into (\ref{eq:stt}), resulting in:
\begin{equation}\label{eqpsi}
    \psi'=-\mathcal{A}\left(\psi-\mathcal{R}_{1}\right)\left(\psi-\mathcal{R}_{2}\right)\psi,
\end{equation}
where
\begin{align}\label{eq:roots}
    \mathcal{A}=
    &
    \frac{l_{\rm P}^{2}r\left[\left(r^{2}+\alpha l_{\rm P}^{2}\right)^{2}+\alpha l_{\rm P}^{4}\right]}{\left(r^{2}+\alpha l_{\rm P}^{2}\right)^{2}\left[r^{2}+(\alpha-1)l_{\rm P}^{2}\right]}, \nonumber\\
    \mathcal{R}_{1,2}=
    &
    -\left[
\vphantom{ \left.\left.+l_{\rm P}^{2}\left[\left(r^{2}+\alpha l_{\rm P}^{2}\right)^{2}-r^{4}/2\right]\right\}\right)^{1/2}}
\left(r^{2}+\alpha l_{\rm P}^{2}\right)^{2}+l_{\rm P}^{2}\left(r^{2}/2+\alpha l_{\rm P}^{2}\right)\right. \nonumber \\     &
    \pm \left(\left[r^{2}+\alpha l_{\rm P}^{2}\right]^{4}-l_{\rm P}^{2}\left\{r^{2}\left(r^{2}+\alpha l_{\rm P}^{2}\right)^{2} \right.\right.     \nonumber\\    
&     
    \left.\left.\left.+~l_{\rm P}^{2}\left[\left(r^{2}/2+\alpha l_{\rm P}^{2}\right)^{2}-r^{4}/2\right]\right\}\right)^{1/2}\right] 
    \nonumber \\ &
    \times \left[\mathcal{A}\left(r^{2}+\alpha l_{\rm P}^{2}\right)\left(r^{2}+(\alpha-1)l_{\rm P}^{2}\right)\right]^{-1}. 
\end{align}
Given this system of equations, \eqref{eq:srr} and \eqref{eqpsi}, there are several useful observations to make.

In an attempt to solve this nonlinear equation \eqref{eqpsi} one notices a strange divergent behavior at \mbox{$r^{2}=(1-\alpha)l_{\rm P}^{2}$}, where the denominator of $\mathcal{A}$ vanishes.
The introduction of the positive parameter $\alpha$ as regulator of the Polyakov RSET is enough to construct a regular Regularized Polyakov RSET for any given fixed background spacetime. However, when dealing with self-consistent solutions that take back reaction into account, we need more stringent conditions. To completely remove divergences caused by an ill-behaved RSET we need to take $\alpha$ greater than 1. Otherwise, we will face a singularity at $r^{2}=(1-\alpha)l_{\rm P}^{2}$ reminiscent of the divergence of the Polyakov RSET at $r=0$. Previous works have treated this singularity as a semiclassical version of the Schwarzschild central singularity \cite{FabbriNavarro2005} or as a numerical instability limiting the resolution of numerical analyses \cite{Ho:2017joh}. 
Our understanding is that the unphysical divergence at $r=0$ of the Polyakov RSET is transformed by the non-linearity of the semiclassical equations into a singularity at $r=\sqrt{1-\alpha}l_{\rm P}$. This displaced singularity cannot be removed by just taking $\alpha>0$: we need to take $\alpha>1$ and we shall proceed in this manner. By removing this previous divergence the solutions to \eqref{eqpsi} can now be explored all the way up to $r=0$ without any restrictions.

Coming back to equation \eqref{eqpsi}, we can see that the right-hand side is written as a cubic polynomial in $\psi$. One can easily check that the non-vanishing roots of this polynomial,
$\mathcal{R}_{1}$ and $\mathcal{R}_{2}$,  are negative definite for any positive value of the radial coordinate and the regulator parameter $\alpha$. The sign of $\psi'$ can be determined by inspection of (\ref{eqpsi}) depending on whether $\psi$ takes values on the different intervals defined by $\mathcal{R}_{1}, \mathcal{R}_{2}$, and $0$, being monotonic within each of these intervals.

One can also easily check that equation \eqref{eqpsi} has two non-trivial exact solutions
\begin{equation}\label{exactsol}
\psi_{\pm}=
   -\frac{r^{2}+\alpha l_{\rm P}^{2}}{r l_{\rm P}^{2}}\left(1\pm\sqrt{1-\frac{l_{\rm P}^{2}}{r^{2}+\alpha l_{\rm P}^{2}}} \right).
\end{equation}
These solutions, when plugged in equation \eqref{eq:srr}, lead to negative infinite values of the compactness, and hence, to an infinite Kretschmann scalar. Therefore these solutions are not physical. Nevertheless, their interest resides in the fact that \eqref{eqpsi} is a first order differential equation that satisfies the hypotheses of Picard-Lindelöf's theorem. Thus, it is guaranteed that no other exact solution will intersect $\psi_{\pm}$ at any finite radius. Additionally, both of the solutions in \eqref{exactsol} are negative for any $r$ and $\alpha$, as are the roots \eqref{eq:roots}.

Another particular feature of the semiclassical equations involves the radial Einstein equation (\ref{eq:srr}), which can be written as a quadratic polynomial in $\psi$. We can solve this quadratic equation to express $\psi$ in terms of $C$:
\begin{equation}\label{eq:psicomp}
\psi=
   -\frac{r^{2}+\alpha l_{\rm P}^{2}}{r l_{\rm P}^{2}}\left(1\pm\sqrt{1+\frac{l_{\rm P}^{2}}{r^{2}+\alpha l_{\rm P}^{2}}\frac{C}{1-C}} \right).
\end{equation}
This expression has two branches depending on the $\pm$ sign. It is interesting to notice
that only the branch with the $-$ sign returns the classical relation \eqref{eq:ClassicalEqs} in the $l_{\rm P}\to0$ limit. The $+$ sign branch does not have a well defined classical limit and therefore is inherently semiclassical. We shall call this branch the concealed branch, and the other one the unconcealed branch. As we will show, the solutions that we are going to describe typically exhibit smooth jumps between the two branches. 

In the next section we shall proceed with the construction and analysis of the solutions to the previous set of equations. We will integrate the system from the asymptotic infinity inwards, and we will mathematically show the qualitative features of the solutions. We will provide approximate analytical expressions in specific local regions and we will also show some numerical integrations.

\section{Vacuum solutions}
\label{Sec:Solution}

Let us start the analysis of the solutions by imposing conditions at the only asymptotically flat region. We provide here a brief summary of the discussion below for the benefit of the reader. In Sec. \ref{Subsec:Schwarzschild} we check that the semiclassical solutions have the expected behavior in the asymptotically flat region (that is, they are equivalent to the Schwarzschild solution with positive ADM mass, up to subleading corrections in the limit $r\rightarrow\infty$), and we also establish the monotonicity of the function $\psi$. This monotonicity is used in Sec. \ref{sec:nonullr} to show that there are no solutions in which $r=0$ is reached; in other words, the domain of definition of $\psi(r)$ must be bounded from below by a certain $r_{\rm B}>0$. Then, we determine the properties of $\psi$ around $r=r_{\rm B}$, showing that the geometry displays a wormhole throat for this value of the radial coordinate. In Sec. \ref{sec:through} we integrate the semiclassical equations on the other side of the wormhole, discussing the relevant metric and asymptotic properties of the portion of spacetime beyond the throat. For completeness, in Sec. \ref{sec:negadm} we discuss solutions with different boundary conditions, namely negative and vanishing ADM mass.

\subsection{Asymptotically flat regime}
\label{Subsec:Schwarzschild}

Let us start by assuming that: i) $C$ is positive at a fiducial reference radius $r_{\text{ref}}$ and ii) we are in the unconcealed branch of $\psi$. Under these two conditions, on the one hand we know that $\psi$ is positive by virtue of Eq. \eqref{eq:psicomp}. This is so because the branch of $\psi$ that has a well defined classical limit guarantees a positive $\psi$ when $C>0$. On the other hand, we are also sure that $\psi$ decreases monotonically towards larger radii, because the roots in the right-hand side of Eq. \eqref{eqpsi} are negative. Now,
$\psi$ cannot cross $\psi=0$ at a finite radius because $\psi=0$ is an exact solution of Eq. \eqref{eqpsi} and so it cannot be intersected by any other solution. In addition, $\psi$ cannot tend to a constant positive value in the limit $r\to\infty$ because in that case $\psi'$ would not go to zero, producing a contradiction. Then, the only remaining possibility is that $\psi$ tends to 0
asymptotically with $r$.

Let us assume a polynomial decay for the asymptotic form of $\psi$:
\begin{equation}
    \psi\propto r^{-\eta}, \quad \eta>0, \quad \text{when } r\to\infty.
\end{equation}
This means that $\psi'$ is proportional to $-\eta\, r^{-\eta-1}$. On the other hand, replacing the previous ansatz in \eqref{eqpsi} returns the relation:
\begin{equation}\label{eq:psieta}
    \psi'\propto-2r^{-\eta-1}+\cdots
\end{equation}
where subdominant terms in $r$ have been neglected.
Therefore, we obtain $\eta=2$, i.e.:
\begin{equation}\label{eq:psiSchwarzschild}
    \psi\simeq\frac{\psi_{0}}{r^{2}}.
\end{equation}
Here, $\psi_{0}$ comes out as an integration constant. Now, the redshift function $\phi$ is obtained by integration of \eqref{eq:psiSchwarzschild} and, in turn, the time component of the metric is:
\begin{equation}
    e^{2\phi}= e^{- {2\psi_{0}}/{r}}\simeq \left(1-\frac{2\psi_{0}}{r}\right),
\end{equation}
where we have got rid of an irrelevant rescaling of time.

Finally, the compactness can be obtained through (\ref{eq:srr}) and for large radii is found to be:
\begin{equation}\label{eq:CompSchwarzschild}
    C\simeq\frac{2\psi_{0}}{r}.
\end{equation}
Fixing the integration constant to be $\psi_{0}=M$ we conclude that the semiclassical counterpart to the Schwarzschild vacuum solution has the same asymptotic properties as the Schwarzschild solution. Consistent with that, we can see that the tangential pressures induced by the Regularized Polyakov RSET vanish in the asymptotic region:
\begin{equation}\label{eq:asympG}
    \langle \hat{T}_{\theta\theta}(r\to\infty)\rangle^{\rm RP}\simeq-\frac{\alpha M^{2} l_{\rm P}^{4}}{8\pi r^{6}}.
\end{equation}

A distant observer should not be able to distinguish any semiclassical departure from classical general relativity. This happens because the density of the quantum substance diminishes towards infinity at a rate greater than $1/r^{2}$ and, in fact, proportional to $1/r^{5}$. It is distributed in such a faint way that, at radial infinity, vacuum polarization does not prevent spacetime from being flat. However, as we show in the following, as we go towards the internal region in our integration, semiclassical deviations from the Schwarzschild metric start taking a prominent role. These deviations become extreme as we get close to $r=2M$, completely removing the horizon.

\subsection{Integrating inwards \label{sec:nonullr}}

The function $\psi$ must be positive and monotonically increasing towards the interior. To determine the qualitative behavior of $\psi$, let us proceed by discarding possibilities. First, let us suppose that $\psi$ goes to a positive constant at $r=0$. Then we can perform the following expansion around $r=0$ in Eq. (\ref{eqpsi}):
\begin{equation}\label{eq:psi0}
    \psi'=-\frac{2\alpha}{(3\alpha-1)}\frac{\psi}{r}+\mathcal{O}(r^{0}).
\end{equation}
Here we can see that such constant value should be reached with an infinite derivative. This is not possible, as can be seen by solving the above differential equation. Integrating Eq. \eqref{eq:psi0} we obtain
\begin{equation}\label{eq:psilocal}
    \psi\simeq r^{-2\alpha/(3\alpha-1)},
\end{equation}
up to a multiplicative integration constant. Since for $\alpha>1$ the above exponent is negative, such a solution would be divergent at $r=0$, thus contradicting our initial assumption that $\psi$ would reach a (positive) constant value at $r=0$.

Next, let us assume that $\psi \to +\infty$ at $r=0$. Depending on the rate at which $\psi$ diverges, various terms can dominate the right-hand side of Eq. \eqref {eqpsi} close to the origin. We have three possibilities depending on the following limit:
\begin{equation}
    \lim_{r \to 0} \psi r = \left\{ 
    \begin{tabular}{c}
         $+\infty$  \\
         $\rm{constant}\neq 0$  \\
         $0$  
    \end{tabular}
    \right.
    .
\end{equation}
For the first case, $\displaystyle{\lim_{r\to0} \psi r} = +\infty$, the differential equation \eqref{eqpsi} acquires the approximate form
\begin{equation}
    \psi'\simeq-\frac{1+\alpha}{(\alpha-1)\alpha}r\psi^{3},
\end{equation}
whose solutions are
\begin{equation}
    \psi\simeq\pm\sqrt{\frac{\alpha(\alpha-1)}{\alpha+1}}r^{-1}.
\end{equation}
Both solutions $\left(\pm\right)$ lead to a contradiction with the initial hypothesis. In the same manner, in the case in which $\displaystyle{\lim_{r\rightarrow0} \psi r} = 0$, we find the following approximate differential equation
\begin{equation}
    \psi'\simeq-\frac{2\alpha}{\alpha-1}\frac{\psi}{r}.
\end{equation}
The solution of this differential equation is
\begin{equation}
    \psi\simeq r^{-2\alpha/(\alpha-1)},
\end{equation}
which, for $\alpha>1$, has an exponent smaller than $-2$, again contradicting the initial hypothesis. Now we pass to the remaining case, $\displaystyle{\lim_{r\to0} \psi r} =\lambda>0$. Then, equation (\ref{eqpsi}) returns at leading order
\begin{equation}
    -\frac{\lambda}{r^{2}}\simeq-\frac{2\alpha^{2}\lambda+2\alpha(1+\alpha)\lambda^{2}+(1+\alpha)\lambda^{3}}{r^{2}(\alpha-1)\alpha},
    \end{equation}
which is satisfied for the values 
\begin{equation}\label{eq:lambda}
     \lambda=-\alpha\pm\sqrt{\alpha(\alpha-1)}.
\end{equation}
Given that these two values are negative, we find again a contradiction with the initial hypothesis. As a consequence, no solutions with positive (finite or divergent) $\psi$ at $r=0$ exist. Therefore the only remaining possibility is that $\psi$ diverges at some finite nonzero radius that we shall call $r_{\rm B}$. 

Let us now analyze the form of this divergence at  $r=r_{\rm B}$. Again, by assuming that $\psi\to+\infty$ when $r\to r_{\rm B}$, we can locally simplify Eq. \eqref{eqpsi}. In fact we can neglect all the powers of $\psi$ less than cubic, thus arriving at the relation
\begin{equation}\label{eq:psidiv}
    \psi'\simeq-\frac{\left[(r^{2}+\alpha l_{\rm P}^{2})^{2}+\alpha l_{\rm P}^{4}\right]l_{\rm P}^{2}r\psi^{3}}{\left[r^{2}+(\alpha-1)l_{\rm P}^{2}\right](r^{2}+\alpha l_{\rm P}^{2})^{2}}.
\end{equation}
The exact solutions to this differential equation are given by
\begin{align}\label{eq:psiapprox}
    \psi=
    &
    \pm l_{\rm P}^{-1}\bigg [ \frac{\alpha l_{\rm P}^{2}\left(r^{2}-r_{\rm B}^{2}\right)}{\left(r^{2}+\alpha l_{\rm P}^{2}\right)\left(r_{\rm B}^{2}+\alpha l_{\rm P}^{2}\right)}-\alpha\ln\frac{r^{2}+\alpha l_{\rm P}^{2}}{r_{\rm B}^{2}+\alpha l_{\rm P}^{2}}\nonumber\\
    &
    +(1+\alpha)\ln\frac{r^{2}+(\alpha-1)l_{\rm P}^{2}}{r_{\rm B}^{2}+(\alpha-1)l_{\rm P}^{2}}
    \left. \vphantom{\frac{\alpha l_{\rm P}^{2}}{r^{2}+\alpha l_{\rm P}^{2}}}
    \right]^{-1/2}.
\end{align}
Only the positive sign in Eq. \eqref{eq:psiapprox} is consistent with our initial hypothesis of asymptotic flatness. Notice that restricting to positive $\psi$ amounts to maintaining the solution in the unconcealed branch of \eqref{eq:psicomp}.

The divergent behavior can be more easily seen by expanding the logarithms in the limit $r\to r_{\rm B}$ up to first order. The solution then acquires the simplified form
\begin{equation}\label{eq:psirb}
    \psi\simeq\sqrt{\frac{k_{0}}{4(r-r_{\rm B})}},
\end{equation}
where the constant
\begin{equation}
    k_{0}=\frac{2\left[r_{\rm B}^{2}+(\alpha-1)l_{\rm P}^{2}\right]\left(r_{\rm B}^{2}+\alpha l_{\rm P}^{2}\right)^{2}}{r_{\rm B}l_{\rm P}^{2}\left[(r_{\rm B}^{2}+\alpha l_{\rm P}^{2})^{2}+\alpha l_{\rm P}^{4}\right]}>0
\end{equation}
has absorbed all dependence on the regulator parameter~$\alpha$.

By integrating Eq. \eqref{eq:psirb} we can deduce the form of the function $\phi$ in a neighborhood of $r_{\rm B}$:
\begin{equation}
    \phi(r)=\phi_{\text{ref}}+ \int_{r_{\text{ref}}}^{r}\psi(r')dr'.
\end{equation}
Owing to the specific divergence of $\psi$, proportional to $(r-r_{\rm B})^{-1/2}$, it follows that $\phi$ does not go to $-\infty$ when $r$ goes to $r_{\rm B}$ (which we can always assume to be smaller than $r_{\text{ref}}$). Specifically, we obtain the form
\begin{equation}
    \phi\simeq\sqrt{k_{0}(r-r_{\rm B})}+\phi_{\rm B}.
\end{equation}
Now the compactness function can be obtained from Eq. \eqref{eq:psirb} up to leading order in $r-r_{\rm B}$:
\begin{equation}
    C\simeq
    1-k_{1}\left(r-r_{\rm B}\right),
    \end{equation}
with
\begin{equation}
    k_{1}=\frac{4\left(r_{\rm B}^{2}+\alpha l_{\rm P}^{2}\right)}{r_{\rm B}^{2}l_{\rm P}^{2}k_{0}}.
    \nonumber
\end{equation}
Since $C$ goes to $1$ as $r\to r_{\rm B}$, it seems that the metric is singular at this radius. However, we can check that this is not the case by changing the radial coordinate from $r$ to a proper radial coordinate $l$, defined through the relation
\begin{equation}\label{eq:ProperCoord}
    \frac{dl}{dr}=\frac{1}{\sqrt{k_{1}(r-r_{\rm B})}}.
\end{equation}
Integrating this definition returns
\begin{equation}\label{eq:PropCoordInt}
    r-r_{\rm B}=\frac{k_{1}}{4}\left(l-l_{\rm B}\right)^{2}.
\end{equation}
Now with this coordinate the resulting metric for $l\gtrsim l_{\rm B}$ can be written as
\begin{align}\label{eq:WormhMetric}
    ds^{2}\simeq
&
    -\exp[\sqrt{k_{0}k_{1}}(l-l_{\rm B})+2\phi_{\rm B}]dt^{2}\nonumber\\
&
    +dl^{2}+\left[\frac{k_{1}}{4}\left(l-l_{\rm B}\right)^{2}+r_{\rm B}\right]^{2}d\Omega^{2}.
\end{align}
This non-singular form of the metric hints to the possibility of extending the geometry beyond $l=l_{\rm B}$. In terms of the radial coordinate $r$ we were using before, this would imply the presence of a second branch in which now $r$ increases as $l$ decreases. The relation between the radial coordinates on this second branch would be
\begin{equation}\label{eq:negRel}
    \frac{dl}{dr}=-\frac{1}{\sqrt{k_{1}(r-r_{\rm B})}}.
\end{equation}
Indeed, we can explicitly check that this extension exists: the metric for $l\lesssim l_{\rm B}$ implies the following form for $\psi$,
\begin{equation}
    \psi \simeq -\sqrt{\frac{k_{0}}{4(r-r_{\rm B})}},
\end{equation}
and this negatively divergent $\psi$ is a solution of the differential equation \eqref{eqpsi}.
That is, the function $\psi$ must make a jump from $+\infty$ to $-\infty$ at $r_{\rm B}$.
The redshift function $e^{2\phi}$, however, goes through this jump in an absolutely smooth fashion, and remains nonzero. 

In this way we show that the semiclassical vacuum solution corresponding to a positive asymptotic mass acquires a surface with minimal radius $r_{\rm B}$ (a minimal surface) that is, therefore, a wormhole throat as defined in Sec. \ref{Sec:Preliminars}. Indeed, the redshift function is different from zero in passing through the wormhole throat, even if $C\rightarrow1$ as $r\rightarrow r_{\rm B}$; hence, no horizon is formed. This wormhole is not mirror-symmetric through the throat, precisely because of the behavior of the redshift function, which is decreasing in passing through the throat. The geometry around the throat could be made symmetric (entailing a discontinuity in the derivative of the redshift function at $l=l_{\rm B}$), by introducing a shell of matter with a SET proportional to $\delta(l-l_{\rm B})$. Here, however, we stick to the strict vacuum solution. It is also interesting to notice that, at the throat, we are also passing smoothly from the unconcealed to the concealed branch in Eq. \eqref{eq:psicomp}. Therefore, we see that semiclassical solutions do not have a well-defined classical limit, being more than just perturbative modifications of the corresponding classical solutions.

The size of the throat of the wormhole, $r_{\rm B}$, can be arbitrarily small. This result is specific to this paper and comes from the presence of the regulator. In the previous work \cite{Fabbrietal2005}, the radius of the throat cannot be smaller than $l_{\rm P}$ owing to the unphysical divergence of the Polyakov RSET. Here, we wanted to check if, by regularizing the RSET, new types of solutions could appear close to $r=0$. However, we have seen that one just obtains a (regularized) extension of the family of wormhole solutions. 

Before ending this section let us comment that Eq. \eqref{eq:WormhMetric} provides a reliable approximation in the following regimes. If $r_{\rm B}\gg\sqrt{\alpha}l_{\rm P}$, then 
\begin{equation}
    0<r-r_{\rm B}\ll\frac{l_{\rm P}^{2}}{r_{\rm B}}.
\end{equation}
On the other hand, if the wormhole throat is small, ~($r_{\rm B}\ll \sqrt{\alpha}l_{\rm P}$), then it must be
 \begin{equation}
    0<r-r_{\rm B}\ll r_{\rm B}.
\end{equation}
Let us now continue integrating the system of equations inwards.

\subsection{Through the wormhole \label{sec:through}}

As a consequence of the disappearance of the classical event horizon, a new region of spacetime emerges. This portion of spacetime where $\psi$ takes negative values (concealed branch) has characteristics very different from that of the unconcealed branch. As we will show, the semiclassical vacuum generates a new internal asymptotic region. The geometry of the other side of the wormhole can be determined from arguments involving Eq. \eqref{eqpsi}, and similarly to previous situations we can also provide an analytic description of the new asymptotic region.

The roots $\mathcal{R}_{1,2}$ and the unphysical exact solutions $\psi_{\pm}$ diverge towards $-\infty$ as $r\to0$. Given that the boundary conditions at the throat imply $\psi\to-\infty$ as $l\to l_{\rm B}^{-}$ \quad($r\to r_{\rm B}$ from the inside), it is guaranteed that $\psi$ will take values below the two roots and the two unphysical exact solutions close enough to the throat.

\begin{figure}
    \centering
    \includegraphics[width=\columnwidth]{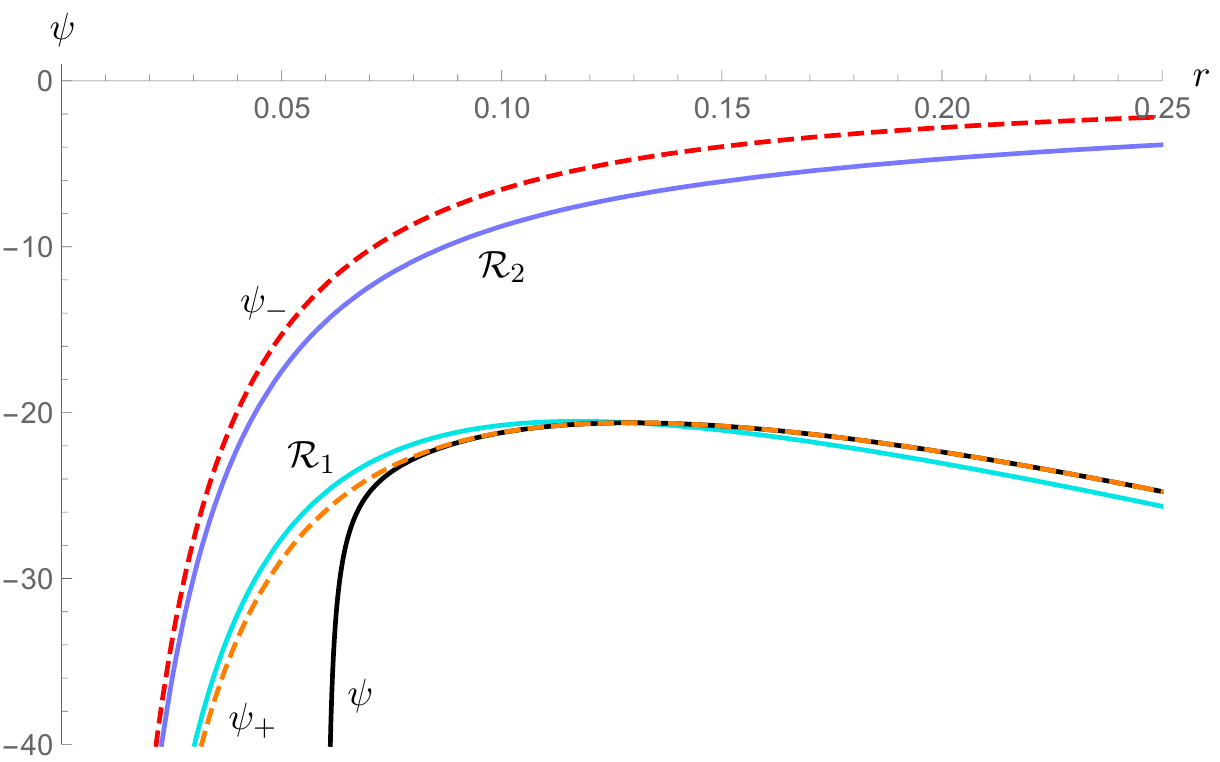}
    \caption{Plot of $\mathcal{R}_{1}$, $\mathcal{R}_{2}$ (cyan and blue curves, respectively) and the unphysical exact solutions, with $\psi_{+}$ being the orange dashed curve and $\psi_{-}$ the red dashed curve for $\alpha=1.01$. The black curve corresponds to a numerical solution with $r_{\rm B}=0.06$. The numerical solution intersects $\mathcal{R}_{1}$ at $r\approx0.13$, reaching a maximum, and then decreases, remaining confined between $\mathcal{R}_{1}$ and $\psi_{+}$. }
    \label{fig:psi_nopert_reg1}
\end{figure}

In figure \ref{fig:psi_nopert_reg1} we have plotted the two roots and unphysical exact solutions from equation \eqref{eqpsi}. We have also plotted a numerical solution with the appropriate behavior at the throat. Let us describe the specific qualitative characteristics of the solution. 
The function $\psi$ has to be monotonically increasing with $r$ (in the decreasing $l$ branch) up to its crossing with $\mathcal{R}_{1}$, something that necessarily takes place. 
Then it starts decreasing but it can neither cross back the root $\mathcal{R}_{1}$ nor cross the exact solution. As both $\mathcal{R}_{1}$ and the unphysical exact solution $\psi_{+}$ have the same asymptotic behavior with $r$, the physical exact solution must acquire this same asymptotic behavior but always living between these two curves. The unphysical solution $\psi_{+}$ acts as an attractor to which solutions converge.

The metric in the asymptotic region can be determined by assuming that $\psi$ deviates slightly from the unphysical exact solution. Parameterizing this deviation by a function $\chi(r)$, such that $\psi= \psi_{+}+\chi(r)$, we can replace this expression in (\ref{eqpsi}) and solve for $\chi$. Performing an asymptotic expansion at $r\to\infty$, keeping only terms linear in $\chi$, and dropping terms decreasing faster than $r^{-3}$ asymptotically, we obtain the following differential equation:
\begin{equation}\label{eq:chi}
    \chi'=-\mathcal{D}\chi+\mathcal{O}\left(\chi^{2}\right),
\end{equation}
where 
\begin{equation}
    \mathcal{D}=\frac{\left[16 r^{4}+8l_{\rm P}^{2} r^{2} \left(2\alpha-1\right)+l_{\rm P}^{4}\left(32\alpha-5\right)\right]}{4l_{\rm P}^{2}r^{3}}\nonumber.
\end{equation}
We can solve Eq. (\ref{eq:chi}) to obtain the deviation from the exact solution, valid in the limit $r\to\infty$,
\begin{equation}\label{eq:chiapprox}
    \chi
    \simeq-\frac{\chi_{0}}{l_{\rm P}}\left(\frac{r}{l_{\rm P}}\right)^{2-4\alpha}e^{- {2r^{2}}/{l_{\rm P}^{2}}}\left[1-\frac{\left(5-32\alpha \right)l_{\rm P}^{2}}{8r^{2}}\right],
\end{equation}
where $\chi_{0}$ is an dimensionless integration constant. This result is consistent (when taking $\alpha=0$) with the approximate behavior found by Ho and Matsuo in \cite{HoMatsuo2017}. The sign of $\chi$ is negative due to the solution $\psi$ approximating $\psi_{+}$ from below. The presence of a regulator causes a faster decay of the deviation (\ref{eq:chiapprox}) for large $r$. Given the convergent behavior of $\psi$ towards the unphysical solution $\psi_{+}$, we can check that the compactness grows exponentially towards minus infinity,
\begin{equation}
    C\simeq-\frac{\left(r/l_{\rm P}\right)^{4\alpha-3}}{2\chi_{0}}e^{ {2r^{2}}/{l_{\rm P}^{2}}}\left[1+\frac{(9-32\alpha)l_{\rm P}^{2}}{8 r^{2}}\right].
\end{equation}
Therefore, we arrive at the following asymptotic form for the line element:
\begin{align}\label{eq:metricnopert}
    ds^{2}
    \simeq
    &
    \left(\frac{r}{l_{\rm P}}\right)^{1-4\alpha}e^{- {2r^{2}}/{l_{\rm P}^{2}}}\left\{-a_{0}\left(1-\frac{l_{\rm P}^{2}}{8 r^{2}}\right)dt^{2}\right. \nonumber\\
    &
    \left.+\frac{2\chi_{0} r^{2}}{l_{\rm P}^{2}}\left[1-\frac{(9-32\alpha)l_{\rm P}^{2}}{8r^{2}}\right]dr^{2}\right\}+r^{2}d\Omega^{2},
\end{align}
where $a_{0}$ is a constant coming from the integration of $\psi$.
As a consequence of the divergent behavior of $C$, the resulting geometry has a null singularity at radial infinity: the time and radial components of the metric vanish. This asymptotic region is singular and leads to a scalar curvature which blows up exponentially towards negative values:
\begin{equation}
R\simeq-\frac{e^{ {2r^{2}}/{l_{\rm P}^{2}}}(2\alpha-1)}{l_{\rm P}^{2}\chi_{0}}\left(\frac{r}{l_{\rm P}}\right)^{-5+4\alpha}.
\end{equation}
This singular region is located at a finite proper distance from the throat. This can be seen by integrating the asymptotic form
\begin{equation}
     \left(\frac{dl}{dr}\right)^{2}=2\chi_{0}\left(\frac{r}{l_{\rm P}}\right)^{3-4\alpha}e^{-{2r^{2}}/{l_{\rm P}^{2}}}\left[1-\frac{(9-32\alpha)l_{\rm P}^{2}}{8r^{2}}\right].
\label{eq:properdistance}
\end{equation}
The exponential factor leads to a finite proper distance $l_{\rm S} < l_{\rm B}$ for the location of this internal asymptotic region.

For the sake of completeness we have calculated the form of the angular components of the Regularized Polyakov RSET in the internal asymptotic region:
\begin{equation}
     \langle \hat{T}_{\theta\theta}(r\to\infty)\rangle^{\rm RP}\simeq -\frac{\alpha e^{2r^{2}/l_{\rm P}^{2}} }{4\pi\chi_{0}}\left(\frac{r}{l_{\rm P}}\right)^{-3+4\alpha}.
\label{eq:angularterms}
\end{equation}
Perhaps surprisingly, this is nonzero in the internal asymptotic region but in fact diverges. However, one has to take into account that this $r\to\infty$ regime is completely different from that on the external asymptotic region. Essentially the divergence of $C(r)$ in the internal asymptotic region compensates any damping factor in $1/r$.

\subsection{Other asymptotic behaviors \label{sec:negadm}}

For completeness, we want to finish this section by describing the remaining solutions to the semiclassical equations: those with a negative and zero asymptotic mass.

Based on our results for the asymptotically flat regime, we first consider the analysis of the geometry when endowed with a negative asymptotic mass. In this case, the functions $\psi$ and $C$, assuming the unconcealed branch, take the following asymptotic form:
\begin{equation}
\psi\simeq\frac{-\alpha+\sqrt{\alpha(\alpha-1)}}{r},\qquad C\simeq-\abs{M} r^{-\frac{\alpha}{\sqrt{\alpha(\alpha-1)}}}.
\end{equation}

Given a referential radius $r_{\text{ref}}$ deep enough in the asymptotic region we can check that the function $\psi(r)$ for $r\in (r_{\text{ref}},+\infty)$ is always larger than the unphysical exact solution $\psi_{-}$ and, in turn, than the two roots, which are always smaller than $\psi_{-}$ (see figure \ref{fig:psi_nopert_reg1}). We can see that this is the case by noticing that the unphysical exact solution $\psi_{-}$ can be obtained from Eq. \eqref{eq:psicomp} by taking the limit $C\to-\infty$, while this solution has a finite negative $C$. Integrating the solution inwards, as it cannot cross the unphysical solution and neither can the roots, the solution is monotonically decreasing. Following the same argument as in subsection \ref{Subsec:Schwarzschild} we show that the solution diverges as $1/r$ in the $r\to0$ limit. In this same limit $C(r)$ tends to $-\infty$. This solution has a curvature singularity at $r=0$ that corresponds to the semiclassical counterpart of the naked singularity of the classical Schwarzschild geometry endowed with a negative asymptotic mass.

Finally, the solution with $M=0$ corresponds to Minkowski spacetime, where the zero point energy of the scalar field can be fully subtracted, and hence it does not contribute to curvature. This solution, for which the RSET vanishes, marks the boundary between positive and negative mass solutions, that is, between wormholes and naked singularities. Since for wormhole solutions the radius of the throat $r_{\rm B}$ is directly related to the asymptotic mass $M$, taking $r_{\rm B}\to0$ corresponds to making the mass vanish, thus recovering Minkowski spacetime in this limit.

\section{Discussion}
\label{Sec: Discussion}

\begin{figure}
    \centering
    \includegraphics[width=0.9\columnwidth]{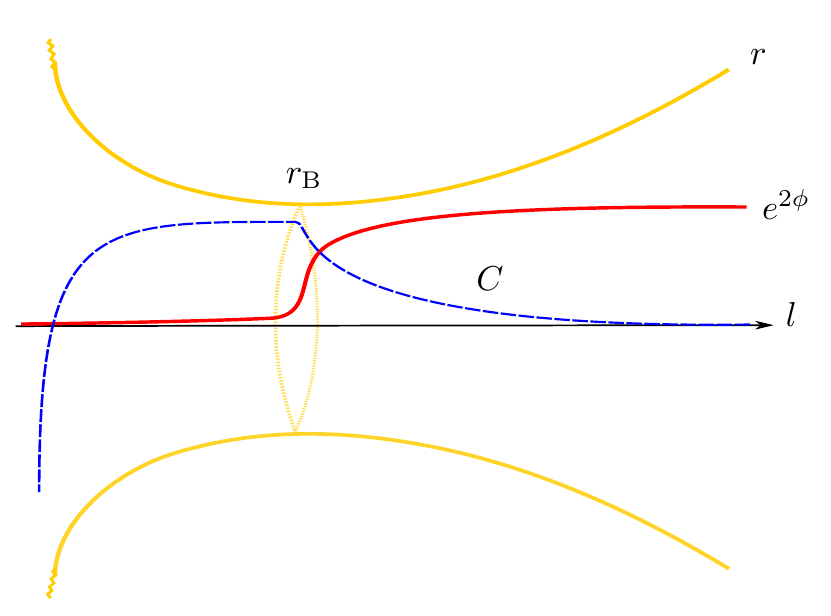}
    \caption{Schematic picture of the semiclassical counterpart of the Schwarzschild vacuum geometry. The horizontal axis is the proper coordinate $l$ while the above and below curves (yellow) represent the radial coordinate $r$. The approximate behavior of the redshift function, in red, and the compactness, in dashed-blue, are drawn over this wormhole geometry. The right side of the wormhole is asymptotically flat whereas the other is asymptotically singular. Both regions are joined by a minimal surface of radius $r=r_{\rm B}$.}
    \label{fig:wormhole}
\end{figure}

\begin{figure}
    \centering
    \includegraphics[width=\columnwidth]{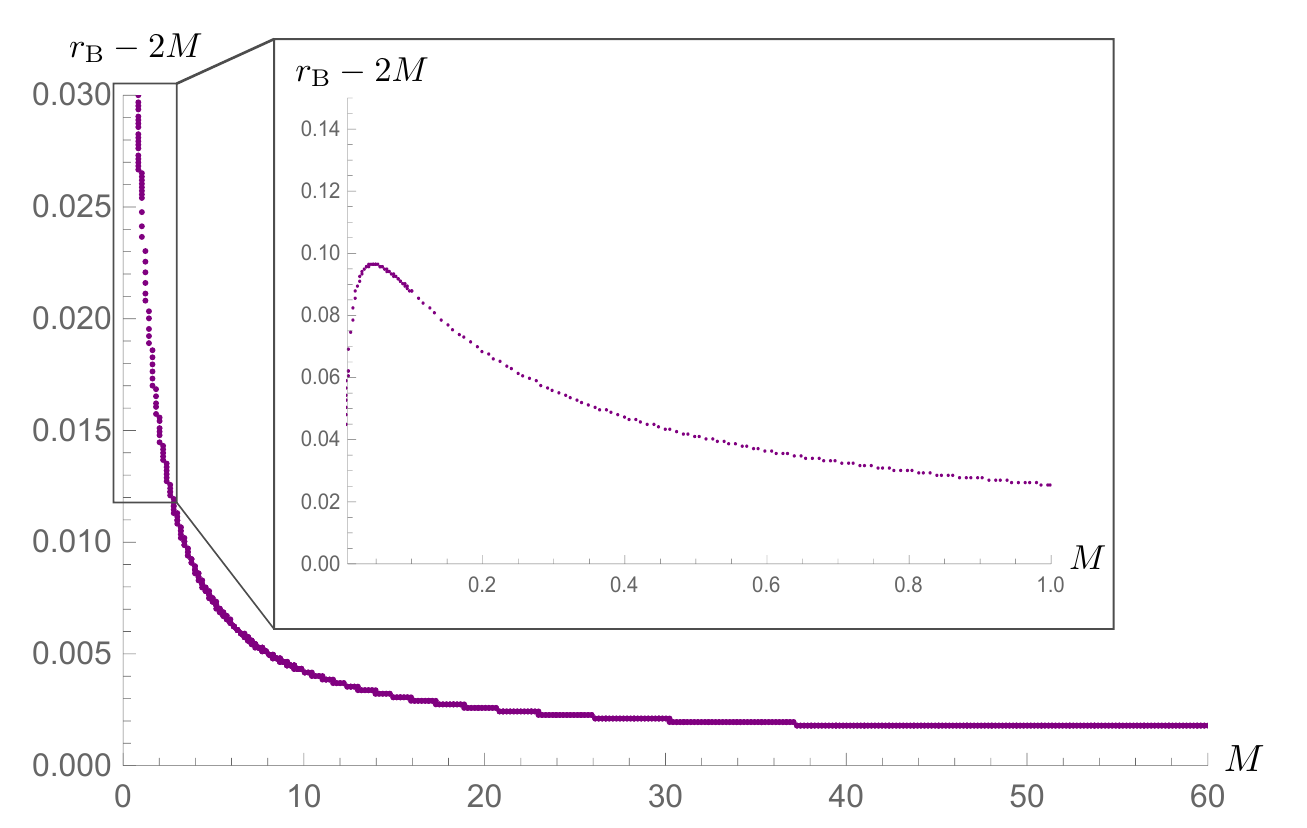}
    \caption{Numerical plot of the deviation $r_{\rm B}-2M$ in terms of the asymptotic mass of the geometry for a range of masses between $10^{-2}$ and $60$. The difference $r_{\rm B}-2M$ reaches a maximum as we approach small values of $M$, while in the $M\to0$ limit $r_{\rm B}$ goes to $0$. For larger masses this quantity is seen to decrease with the mass.}
    \label{fig:RbMasa}
\end{figure}

We have found that the semiclassical vacuum counterpart of the classical positive-mass Schwarzschild solution is an asymmetric wormhole with a singular internal asymptotic region (see figure \ref{fig:wormhole}). In this qualitative plot we can appreciate several interesting features. The external asymptotic region (right-hand side of the picture) is asymptotically flat, while the internal asymptotic region is singular, in the sense that curvature invariants diverge. We have illustrated the asymmetry of the configuration by using the proper radial coordinate $l$.
On the one hand, the compactness function $C$ (dashed line) grows to $1$ at the throat of the wormhole and then decreases towards $-\infty$ in the internal asymptotic region. In terms of the Misner-Sharp mass, $m(r)$ grows from its asymptotic value $M$ up to a value at the throat given by $r_{\rm B}/2$. Then, it starts decreasing reaching $-\infty$ at the internal asymptotic region. In figure \ref{fig:RbMasa} we plot the value of $r_{\rm B}-2M$ with respect to the asymptotic mass $M$ (in Planck units). We see that for a range of masses large enough as compared with the Planck mass, the difference is of the order of $10^{-2}l_{\rm P}$. This difference increases for smaller masses and finally goes to zero for $M \to 0$. Owing to a numerical instability in our method, we have not been able to extend this behavior to larger mass values.
On the other hand, the redshift function is always monotonically decreasing and it only tends to zero at the internal asymptotic region. So we can see that the location where the classical horizon would have been placed is now substituted by a wormhole throat with a nonzero redshift value. In a sense, the horizon has been pushed away towards an internal singular infinity.

The distribution of the Misner-Sharp mass along the radial direction allows for an interesting interpretation. It is as if an infinite negative energy was concentrated in the internal singular region and then there was a cloud of negative vacuum energy distributed throughout the entire spacetime adding to this contribution. In going from the internal asymptotic region towards the throat, this negative semiclassical energy increases the value of the Misner-Sharp mass. This counter-intuitive behavior happens because of the negative relation between $dl$ and $dr$ in Eq. \eqref{eq:negRel}. When one reaches the throat itself the Misner-Sharp mass is already positive. Once the throat has been surpassed, the semiclassical negative energy now progressively decreases the value of the Misner-Sharp mass, leading finally to the asymptotic mass $M$.

\begin{figure}
    \centering
    \includegraphics[width=0.8\columnwidth]{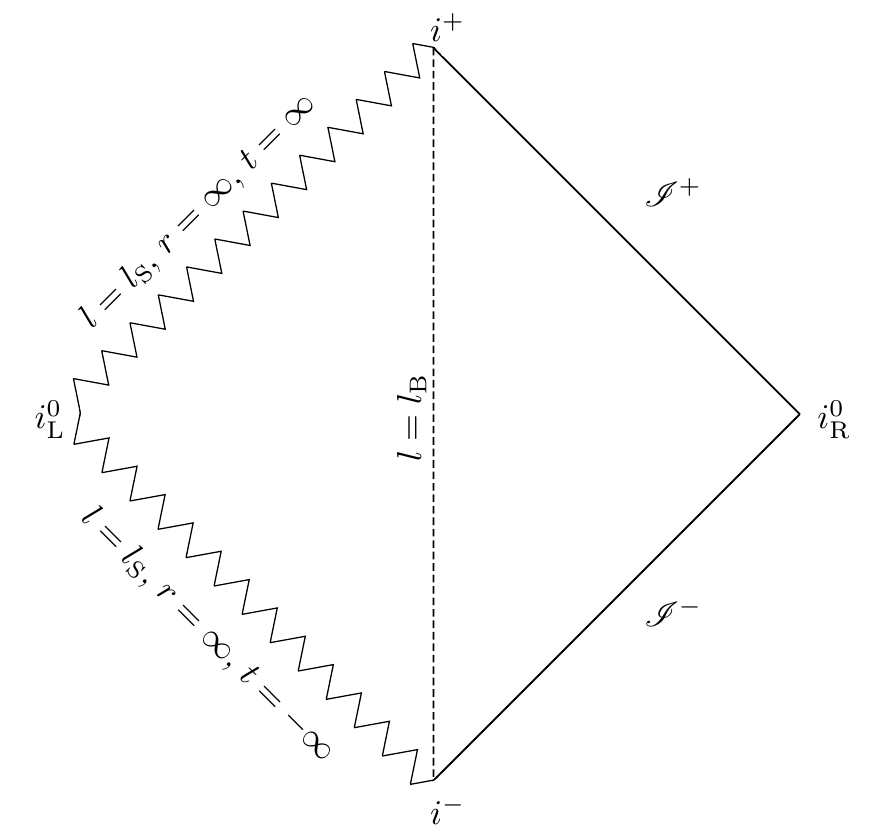}
    \caption{Penrose diagram corresponding to a singular wormhole solution. The vertical dashed line denotes the location of the wormhole neck. To its right, the asymptotically flat portion of spacetime is depicted alongside its asymptotic regions. The left hand side of the diagram shows the internal past and future null singularities, which are located at finite proper distance from the neck $l_{\rm B}-l_{\rm S}$. The point $i^{0}_{\rm L}$ is singular as well, and is reached in finite proper time by spacelike geodesics.}
    \label{fig:penrose}
\end{figure}

Notice that within the vacuum solutions analyzed here there is none which is regular, with the exception of the Minkowskian solution. Moreover, by analyzing the causality of the wormhole solutions (figure \ref{fig:penrose}), one can check that the singular region is null as opposed to the situation in the Schwarzschild solution because of the asymptotic vanishing of the time and radial components of the metric ~\eqref{eq:WormhMetric}. Moreover, observers following timelike trajectories reach this singularity at finite proper time and null rays reach it at a finite value of the affine parameter. Hence, any Cauchy surface would touch the singular region: it is a naked singularity. Such an object could be observationally distinguishable from a Schwarzschild black hole \cite{VirbhadraEllis2002} through their gravitational lensing imprint. Absence of regular self-consistent configurations indicates that there are no ``mass without mass" solutions of any sort (using Wheeler's terminology \cite{Wheeler1955}), in the semiclassical theory: vacuum energy cannot by itself generate regular self-gravitating configurations. The introduction of a material content of some sort is therefore a necessary requirement in order to obtain regular and semiclassically consistent geometries. This situation will be discussed extensively in a forthcoming publication. 
In a similar manner to that in the classical Schwarzschild geometry, the sign of the asymptotic mass $M$ determines how the compactness function $C$ behaves inwards. Negative asymptotic masses make the compactness diverge towards $-\infty$ as $r\to0$. This situation is qualitatively similar to the classical case of a negative mass Schwarzschild geometry. However, for positive asymptotic mass, instead of finding a positive divergence at $r=0$, in the semiclassical case we find that the compactness reaches a minimal surface where $C=1$, and then again diverges negatively but this time in the internal asymptotic region at $r\to+\infty$. The solution corresponding to Minkowski spacetime marks the boundary between these two behaviors, having $C(r)=0$ identically.
In order to analyze potentially regular geometries sourced by an internal matter content plus the semiclassical vacuum contributions it is necessary to have a control of the effects of the semiclassical energies up to arbitrarily small radii. This is the reason why we decided to start analyzing in this paper the semiclassical vacuum solutions with a Regularized Polyakov RSET. We did not know a priori how the regularizing parameter $\alpha$ would affect the solutions found in~\cite{Fabbrietal2005}. Which type of solutions would appear when trying to explore radii arbitrarily close to zero? Would the internal region be affected by the regularization? We have seen that in fact the new regular terms do not disappear in the internal region as $r\to+\infty$ [recall Eq. (\ref{eq:angularterms})] but instead blow up there. However, this behavior does not modify the singular nature of this region; the only effect has been to bring this singular region closer to the throat, in terms of proper distance, than when no regulator is present ($\alpha=0$). This can be seen easily by noticing the factor $1/r^{4\alpha}$ in Eq. (\ref{eq:properdistance}). On the other hand, within the regularized theory there exist solutions that can come as close as desired to $r=0$. No matter how small the asymptotic mass of the system is, the solutions are always wormholes. The Minkowski solution is a singular limit of the solutions as $r_{\rm B} \to 0$.     

Another interesting fact of the semiclassical static solutions that we want to emphasize is the impossibility for these solutions to posses a non-extremal horizon (meaning horizon with a non-zero surface gravity), independently of whether they include some classical matter or not (similar conclusions, but in a different approximation, have been reached in \cite{Berthiere2017}). This can be shown by the following argument. Any self-consistent static semiclassical solution with a non-extremal horizon will have a semiclassical RSET divergent at the horizon. This divergence can be seen directly from Eq. (\ref{eq:}) by calculating e.g. the physical density
\begin{equation}
   \rho=e^{-2\phi} \langle \hat{T}_{tt} \rangle^{\rm P}=
\frac{l_{\rm P}^{2}}{8\pi r^{2}}\left[2\psi'(1-C)+\psi^{2}(1-C)-\psi C'\right].
\label{eq:rho}
\end{equation}
A behavior 
\begin{equation}
   e^{2\phi} \propto (r-r_{\rm H}), \qquad
   1-C \propto (r-r_{\rm H}),
\end{equation}
where $r_{\rm H}$ is the radius of the horizon,
leads to a divergence when plugged into \eqref{eq:rho}. To have a non-extremal horizon this semiclassical divergence would have to be compensated by an equivalent divergence in a classical source. But, in principle,  this is an undesirable procedure.

In view of this, in the semiclassical paradigm, if an equilibrium state is reached at some point during evolution, it should be given by horizonless or extremal configurations.
The standard black hole paradigm circumvents this situation because trapping horizons should be formed dynamically by a collapse process and then start evaporating. Then, it is assumed that the geometry would be never static, except perhaps at the end of a long evaporation process when the horizon itself might disappear. 
Here we are just remarking that, within semiclassical gravity, we can in principle have both: non-equilibrium
configurations which in turn can have non-extremal trapping horizons, on the one hand,
or equilibrium configurations with extremal horizons or no horizons whatsoever (as has been shown to happen in the vacuum solutions), on the other.

For completeness, let us also mention that to preserve a non-extremal event horizon in a static configuration after backreaction has been taken into account, we could make use of the Hartle-Hawking vacuum state. A Hartle-Hawking vacuum state can compensate the divergence of the Boulware vacuum with the inclusion of outgoing and ingoing thermal fluxes of energy. As a consequence, in principle the horizon could be preserved in a self-consistent calculation. However, this has the price of requiring the black hole to be immersed in a thermal bath at precisely the temperature $T=\kappa/2\pi$, with $\kappa$ the surface gravity of the resulting self-consistent black hole. If the incoming thermal flux were not perfectly adjusted, the Boulware divergence would not be exactly canceled and again, the horizon would be destroyed.

\section{Summary} 

In the present work we have studied the set of static vacuum solutions to the semiclassical Einstein field equations sourced by an approximation to the RSET which is regular at $r=0$. Manageable approximations to the RSET are typically divergent there, which compromises any analysis of regular stellar configurations. To circumvent this problem we have prescribed a form of Regularized Polyakov RSET which is at the same time manageable and regular. The semiclassical solutions found in this manner extend and generalize those in~\cite{Fabbrietal2005,HoMatsuo2017}. For the Boulware vacuum state we find that the semiclassical counterpart of the positive-mass Schwarzschild solution is an asymmetric wormhole with a singularity located at a finite proper distance from the throat. The size of the throat can be made arbitrarily small. The semiclassical equations do not have a horizon. It is replaced by a wormhole throat. The semiclassical counterpart to the negative-mass Schwarzschild solution is however qualitatively similar to it, exhibiting a naked singularity. In between these two sets of solutions we trivially find the semiclassically self-consistent Minkowski vacuum.  

Understanding completely the vacuum solutions and having a Regularized Polyakov RSET is an important step for the analysis of the more physical solutions in which there is also a classical matter source in the interior.


\acknowledgments

Financial support was provided by the Spanish Government through the projects FIS2017-86497-C2-1-P, FIS2017-86497-C2-2-P (with FEDER contribution), FIS2016-78859-P (AEI/FEDER,UE), and by the Junta de Andaluc\'ia through the project FQM219. CB and JA acknowledge financial support from the State Agency for
Research of the Spanish MCIU through the ``Center of Excellence Severo Ochoa'' award to the Instituto de Astrof\'{\i}sica de Andaluc\'{\i}a (SEV-2017-0709). RCR acknowledges support from the Preeminent Postdoctoral Program (P$^3$) at UCF.

\bibliographystyle{unsrt}
\bibliography{biblio3}
\end{document}